# Hydrogen Ordering and New Polymorph of Layered Perovskite Oxyhydrides: $Sr_2VO_{4-x}H_x$

Joonho Bang,[1] Satoru Matsuishi,[2] Haruhiro Hiraka,[4] Fumika Fujisaki,[5] Toshiya Otomo,[4,5] Sachiko Maki,[2] Jun-ichi Yamaura,[2] Reiji Kumai,[4,5] Youichi Murakami,[4,5] and Hideo Hosono[*,1,2,3]

[1]*Materials and Structures Laboratory,* [2]*Materials Research Center for Element Strategy, and* [3]*Frontier Research Center, Tokyo Institute of Technology, Yokohama 226-8503, Japan*

[4]*Institute of Materials Structure Science, High Energy Accelerator Research Organization (KEK), Tsukuba, Ibaraki 305-0801, Japan*

[5]*Department of Materials Structure Science, The Graduate University for Advanced Studies, Tsukuba, Ibaraki 305-0801, Japan*

## Abstract

Compositionally tunable vanadium oxyhydrides $Sr_2VO_{4-x}H_x$ ($0 \leq x \leq 1.01$) without considerable anion vacancy were synthesized by high-pressure solid state reaction. The crystal structures and their properties were characterized by powder neutron diffraction, synchrotron X-ray diffraction, thermal desorption spectroscopy, and first-principles density functional theory (DFT) calculations. The hydrogen anions selectively replaced equatorial oxygen sites in the $VO_6$ layers via statistical substitution of hydrogen in the low $x$ region ($x < 0.2$). A new orthorhombic phase (*Immm*) with an almost entirely hydrogen-ordered structure formed from the $K_2NiF_4$-type tetragonal phase with $x > 0.7$. Based on the DFT calculations, the degree of oxygen/hydrogen anion ordering is strongly correlated with the bonding interaction between vanadium and the ligands.

*Corresponding author. E-mail: hosono@msl.titech.ac.jp

The design of functional materials is achieved based on a detailed understanding of the relationship between the material's properties and its atomic structure. In transition-metal oxides, bonding interaction between p orbitals of oxygen anion and d orbitals of transition-metal cation plays an important role in the physical properties of the materials.[1] At this point, exploration of mixed-anion compounds such as oxyhydrides is a promising way of discovering families of functional transition-metal compounds.[2] The bipolar element hydrogen can have valence states from −1 to +1. In addition, the spatial spread of the H 1s orbital is sensitive to the chemical environment.[3] These unique characteristics of hydrogen are expected to provide a new pathway for mediating electronic and magnetic interactions.[2,4] Although several transition-metal oxyhydride compounds have been reported to date, the well-controlled synthesis of these compounds is rather difficult because of opposite chemical nature of hydrogen and oxygen.[2,5–7] We expect two roles for partial replacement oxygen sites by hydrogen in oxides. One is to serve as an effective electron donor. A clear-cut example utilizing this role was recently reported in iron-base superconductors $LaFeAsO_{1-x}H_x$ (where $0 \le x \le 0.5$).[8,9] The other is to work as a unique ligand to a transition-metal cation because an orbital available for $H^-$ is limited to 1s which rather differs from the case of $O^{2-}$.

$Sr_2VO_4$ we focus here is a representative layered perovskite oxide and has a tetragonal $K_2NiF_4$-type structure (space group: $I4/mmm$)[10,11] composed of two-dimensional networks of corner-shared nondistorted $VO_6$ octahedra. This framework is the intergrowth of rock-salt $A$O block and perovskite $ABO_3$ unit, and the structures are well-known to exhibit various intriguing physical phenomena such as high $T_c$ superconductor $La_{2-x}Sr_xCuO_4$. This compound is a Mott insulator with a $d^1$ electronic configuration, and has an antiferromagnetic orbital-ordering transition around 97 K.[12] Here, we report an investigation of the crystal structures and solid-state properties of $Sr_2VO_{4-x}H_x$ ($0.0 \le x \le 1.01$) with controlling of the hydrogen/oxygen amounts in a coordination sphere of vanadium metal cation.

Oxygen/hydrogen ordering was observed by substitution, and a new orthorhombic polymorph was uncovered in highly hydrogen-substituted regions. Incorporation of controllable amount of hydrogen anion to the oxygen site without considerable oxygen-vacancy makes it possible to tune the valence state of vanadium cation from $d^1$ ($V^{4+}$) to $d^2$ ($V^{3+}$) by electron doping. Moreover, the structural stabilization can be expected by switching the HOMO level from the (dπ–pπ)* antibonding molecular orbital to a lower energy dσ–sσ bonding orbital because hydrogen anion has no other orbital except 1s.

The oxyhydrides $LaSrCoO_3H_{0.7}$ and $ATiO_{3-x}H_x$ (A = Ca, Sr, and Ba) were synthesized by a topochemical reaction using reducing agents of NaH or $CaH_2$.[2,7] In the present study, we applied the advanced high hydrogen pressure method to synthesize oxyhydrides, which enables control of the hydrogen content and surpasses the substitution limit of the topochemical method.[13,14] Polycrystalline samples of $Sr_2VO_{4-x}H_x$ in the form of sintered pellets were synthesized by solid-state reaction using a belt-type high-pressure anvil cell, $(2-x)SrO + VO_2 + xSrH_2 \rightarrow Sr_2VO_{4-x}H_x + 0.5xH_2$. $SrH_2$ was prepared by heating strontium metal in a hydrogen atmosphere. All precursors were mixed in a glovebox filled with purified argon gas ($H_2O$, $O_2$ < 1 ppm). The starting mixture was placed in a BN capsule and sandwiched between pellets of external hydrogen source. The external hydrogen source was a mixture of $NaBH_4$ + $Ca(OH)_2$, which releases hydrogen gas during synthesis. The solid-state reaction was performed at 1473 K and 5 GPa for 30 min. BN separators blocked the passage of all reaction products except hydrogen gas. Both the sample and the hydrogen source were enclosed in a NaCl capsule (a hydrogen sealant) that was surrounded by a graphite tube heater and pyrophyllite as a solid pressure-transmitting medium.

Each sample was characterized by powder X-ray diffraction (XRD) using a Bruker D8 Advance diffractometer with Cu $K_\alpha$ radiation. Neutron powder diffraction (NPD) was performed using a neutron total scattering spectrometer (NOVA) installed in the Japan Proton Accelerator Research Complex (J-PARC). The powder diffraction data were measured at room temperature for ~8 h in a vanadium–nickel alloy holder with a diameter of 6 mm. Synchrotron powder X-ray diffraction measurements were performed using the curved imaging plate diffractometer (BL-8A) at the Photon Factory in High Energy Accelerator Research Organization (KEK-PF). The data were analyzed by the Rietveld method using TOPAS,[15] GSAS,[16,17] and RIETAN-FP.[18] The hydrogen concentrations in the samples were analyzed by thermal desorption spectroscopy (TDS) measurements (TDS1200, ESCO). The TDS measurements were carried out in a vacuum chamber with a background pressure of ~$10^{-7}$ Pa at various temperatures from room temperature to 1273 K at a heating rate of 60 K/min. To examine the structural stability, density functional theory (DFT) calculations were performed using the generalized gradient approximation with the Perdew–Burke–Ernzerhof functional[19,20] and the projected augmented plane-wave method[21] implemented in the Vienna ab initio simulation program code (VASP).[22] A $2a \times 2a \times c$ supercell containing nine chemical formulas was used, and the plane-wave basis set cutoff was set to 800 eV. The lattice parameters of each cell were specified using linear fitting of lattice parameters collected by Rietveld fitting of the NPD data. To calculate the total energy, $9 \times 9 \times 6$ $k$-point grids were used.

Because the synthesized $Sr_2VO_{4-x}H_x$ samples were stable in an ambient atmosphere, we performed their characterization in ambient conditions. Figure 1a shows the lattice parameters $a$, $b$, and $c$ of $Sr_2VO_{4-x}H_x$ collected by XRD patterns with different nominal $x$ values ($x_{nom}$) in the starting mixture. For $x_{nom} \leq 0.7$, the XRD patterns were indexed by the $I4/mmm$ tetragonal phase. We found that there were linear relationships between the lattice parameters and $x_{nom}$. $Sr_2VO_4$ crystallizes in a tetragonal phase with a $c/a$ ratio of 3.27, which increases to 3.34 (+2.14%) for $x_{nom}$ = 0.7. At $x_{nom}$ = 0.8, a tetragonal (*I*4/*mmm*)-to-orthorhombic (*Immm*) phase change was observed with a large difference between *a* and *b* ($b - a$ = 0.2037 Å at $x_{nom}$ = 1.2). For $x_{nom}$ > 1.2, variation in lattice constants was saturated.

NPD, which is suited for the investigation of the oxygen and hydrogen sites in crystals, was performed to determine the detailed crystal structures. The NPD patterns for $x_{nom}$ = 0.0, 0.5, and 1.2 are shown in Figure 1b. The crystal of $Sr_2VO_4$ ($x_{nom}$ = 0.0) has vacancies in 2% of the oxygen sites, which were randomly located at both the equatorial (O1) and apical (O2) oxygen positions in the $VO_2$ layers. The crystal structures and local geometries of the $Sr_2VO_{4-x}H_x$ derived from the NPD data are shown in Figure 2. When $x_{nom}$ = 0.5 of hydrogen was introduced into the crystal lattice, the V–O1 length decreased by 0.52%, but the V–O2 length increased by 1.11% with increase of the *c*/*a* ratio. The bond length ratios between V–O1 and V–O2, *d*(V–O2)/*d*(V–O1), were 1.030 and 1.047 for $x_{nom}$ = 0.0 and 0.5, respectively.

Figure 3a shows the TDS data of the $Sr_2VO_{4-x}H_x$ samples with various hydrogen contents. As $x_{nom}$ was increased from 0.3 to 1.4, the amount of hydrogen released as $H_2$ gradually increased from 0.25 to 1.24 per formula unit. We also found that there was broadening of the desorption range and an increase in the number of desorption peaks, which is probably related to the local environment of each hydrogen anion. Considering that the occupation of hydrogen in the O1 site, which has a shorter bond length with V than hydrogen in the O2 site, gradually increased with increasing $x_{meas}$, it appears to be reasonable that the upper limit of the desorption temperature gradually increased in the TDS data. Figure 3b shows the amounts of hydrogen and oxygen vacancies in the samples as a function of $x_{nom}$. The amount of hydrogen was obtained by integrating the thermal desorption spectrum, and the amount of oxygen vacancies was estimated by refining the synchrotron X-ray data (Figure S1). It was found that the amount of hydrogen in the sample ($x_{meas}$) linearly increased with increasing $x_{nom}$.

For $x_{nom}$ = 0.5 and 1.2, we performed Rietveld refinement of NPD data under constraints consistent with the hydrogen contents obtained by TDS. As a result, the compositions of $Sr_2VO_{3.62}H_{0.38}$ and $Sr_2VO_{2.99}H_{1.01}$ were obtained for $x_{nom}$ = 0.5 and 1.2, respectively. We found that the residual factor of refinement $R_{wp}$ value decreased monotonically with a decrease in anion site vacancy when $R_{wp}$ was examined as a function of vacancy fraction for each anion site (Figure S2). This finding means that the vacancy-free model was appropriate for $Sr_2VO_{4-x}H_x$, and this result is in contrast with known layered oxyhydrides such as $LaSrCoO_3H_{0.7}$ and $Sr_3Co_2O_{4.33}H_{0.84}$, where anion vacancies play an important role in stabilizing the crystal structure.[5] The present experimental finding appears to be related to the different synthetic route, i.e., under high-pressure synthetic conditions, oxygen vacancies can be easily filled by $H^-$ owing to combined effects of the high hydrogen pressure in the closed system and stronger covalent nature between V-H which results in the stabilization of hydrogen anion than oxygen vacancy.

We also found that hydrogen in $Sr_2VO_{3.62}H_{0.38}$ shows site selectivity even though the structural phase transition had not yet occurred. In mixed anion compounds with layered structures, there was a strong tendency for isolation of different anions in each site. Therefore, sharing of structurally indistinct sites in the same plane by hydrogen and oxygen is difficult.[5] However, for $Sr_2VO_{3.62}H_{0.38}$, 79% of the hydrogen anions were located at the O1 site, and the O2 site was occupied by 21% of hydrogen anions coexisting with oxygen anions i.e., a mixed-anion state. For $Sr_2VO_{2.99}H_{1.01}$, 97% of the hydrogen anions were located in the V–O plane, and stripe-type ordering of oxygen (O1) and hydrogen (H1) occurred along with the appearance of a new phase by shortening of the V–H1 bond length. Our high-pressure synthesis method with an excess of hydrogen realizes a wide range of hydrogen substitution into oxygen sites without considerable oxygen vacancy.

Figure 4a shows the degree of anion ordering as a function of $x_{\mathrm{meas}}$. In the low hydrogen substitution region ($x_{\mathrm{meas}} < 0.25$), hydrogen anions statistically occupy both O1 and O2 sites. With higher hydrogen substitution ($x_{\mathrm{meas}} > 0.25$), hydrogen anions tend to occupy the oxygen sites in the V–O planes (i.e., the O1 site). Finally, most of hydrogen anions occupy sites in the V–O planes, and a crystallographic phase transition occurs. Figure 4b shows the calculated total energy difference ($\Delta E$) between $E_1$ for H in the O1 site and $E_2$ for H in the O2 site. In the first stage of $x$, $\Delta E$ is so small that anion ordering does not increase the structural stability. When $x > 0.25$, $\Delta E$ starts to decrease, indicating that the system becomes relatively more stable when hydrogen is located in the O1 site. When $x \sim 0.25$, a disorder-to-order transition of hydrogen occurs, which is consistent with calculations of the total energy.

A major difference between $H^-$ and $O^{2-}$ is their electronic structures. Hydrogen has only a 1s orbital available for chemical bonding, whereas the oxygen anion has three available 2p orbitals. As shown in Figure 5, $O^{2-}$ behaves as a $\pi$-donor ligand and its p orbitals have net overlap with the V $d_{xy}$, $d_{xz}$, and $d_{yz}$ orbitals, forming bonding ($\pi$) and antibonding ($\pi^*$) molecular orbitals. The electrons supplied by the oxygen occupy the d$\pi$ p$\pi$ bonding orbitals, whereas an electron originally belonging to the triply degenerate V 3d orbitals ($d_{xy}$, $d_{xz}$, and $d_{yz}$) occupies the antibonding orbital. In contrast, the hydrogen anion has no orbitals that have the same symmetry as the V $d_{xy}$, $d_{xz}$, and $d_{yz}$ orbitals, so the V $d_{xy}$, $d_{xz}$, and $d_{yz}$ orbitals remain nonbonding and are fully localized on the vanadium cation. However, the $H^-$ anion can form strong $\sigma$-bonding with doubly degenerate V 3d orbitals ($d_{x^2-y^2}$, $d_{z^2}$). Therefore, when $O^{2-}$ is substituted by $H^-$, the occupancy in the antibonding orbital decreases, and consequently the total energy of the crystal decreases. The extent of this stabilization should depend on the local environment of hydrogen. In the case of the O1 site, oxygen is coordinated to six cations, 4Sr + 2V, while in the O2 site oxygen is coordinated to 5Sr + 1V. Therefore, when hydrogen is introduced into the O1 site, the stabilization effect is much greater than when it is introduced in the O2 site. The hydrogen substitution-induced phase transition results in the sudden change of the bonding interactions. Because the V–O1 bond length gradually decreases with increasing $x$, the antibonding interaction between V and O1 increases. This structural instability causes elongation of the V–O bonds by the structural transition.

In summary, we synthesized new vanadium-based layered perovskite oxyhydrides $Sr_2VO_{4-x}H_x$, and the hydrogen and oxygen concentrations were successfully controlled in the full range ($0.0 \leq x \leq 1.01$) without any considerable anion-vacancies. This result makes it possible for application of a hydride ion as an effective carrier dopant in oxides and open up the new method for tuning exchange interaction between the transition-metal cations. Moreover, by controlling the hydrogen amount, mixed anion region was uncovered, and it was found that the hydrogen ordering induced structural transition and structural stabilizing effect of hydrogen anion substitution. The theoretical calculations suggest that the degree of oxygen/hydrogen anion ordering is strongly correlated with reduction of antibonding interactions between vanadium and the oxygen ligands. Because a dramatic change is expected in the physical properties with hydrogen substitution so as to drive the change of electronic states, more detailed studies on the physical properties of these compounds are currently underway.

## Acknowledgement

We acknowledge Prof. H. Mizoguchi, Dr. K. Lee, Dr. H. Lei, and Dr. Y. Muraba for valuable discussions. This study was supported by MEXT, Element Strategy Initiative to form a research core. A part of this work was supported by the FIRST program, JSPS.

**Figures Captions**

**Figure 1.** Structural data from the diffraction patterns of $Sr_2VO_{4-x}H_x$. (a) Lattice parameters from the XRD data as a function of $x_{nom}$. (b) NPD patterns with Rietveld fitting of $Sr_2VO_{4-x}H_x$ ($x_{nom}$ = 0.0, 0.5, and 1.2). The data were collected using NOVA (90° bank) at 300 K.

**Figure 2.** Crystal structure of $Sr_2VO_{4-x}H_x$ ($x_{nom}$ = 0.0, 0.5, and 1.2) and local geometry around vanadium cation. Structure analyses from NPD data indicate compositions of $Sr_2VO_{3.62}H_{0.38}$ and $Sr_2VO_{2.99}H_{1.01}$ for $x_{nom}$ = 0.5 and 1.2, respectively. 79% ($Sr_2VO_{3.62}H_{0.38}$) and 97% ($Sr_2VO_{2.99}H_{1.01}$) of hydrogen anions are ordered in V–O planes.

**Figure 3.** Thermal stability and content of hydrogen anions incorporated in $Sr_2VO_{4-x}H_x$. (a) TDS corresponding to the $H_2$ molecule ($m/z$ = 2). (b) Hydrogen content measured by TDS data and oxygen vacancy estimated from synchrotron XRD data as a function of $x_{nom}$.

**Figure 4.** Anion order of $Sr_2VO_{4-x}H_x$ as a function of the hydrogen substitution to oxygen sites. (a) Amount of hydrogen positioned in each plane derived by NPD and synchrotron XRD data. (b) Calculated total energy difference ($\Delta E = E_1 - E_2$) when hydrogen atoms occupy O1 ($E_1$) and O2 ($E_2$) sites in tetragonal $Sr_2VO_{4-x}H_x$.

**Figure 5.** Schematic representation of molecular orbital diagram between vanadium cation and ligand anions (O, H).

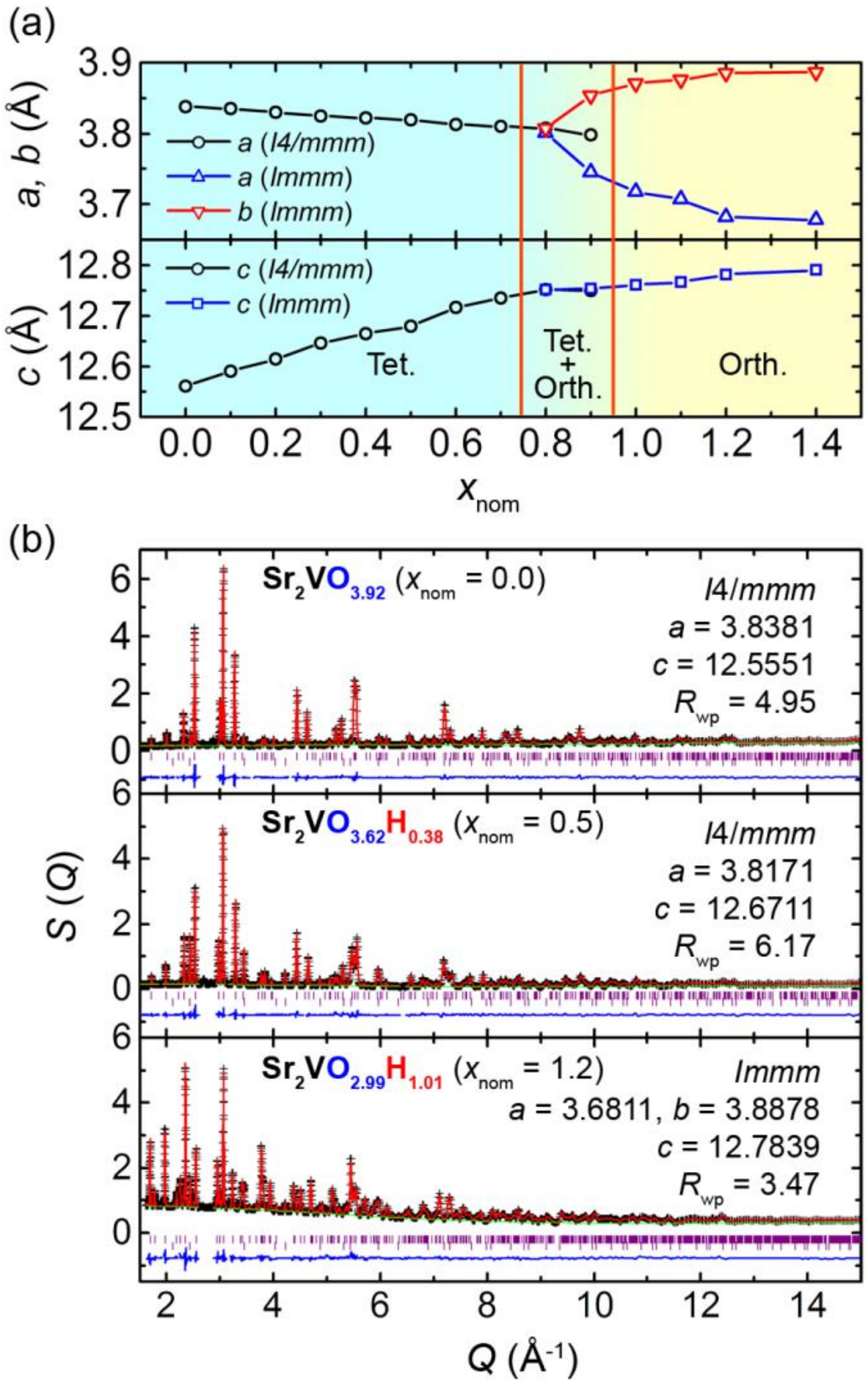
(a)
3.9
3.8
3.7
a, b (Å)
a (I4/mmm)
a (Immm)
b (Immm)
12.8
12.7
12.6
12.5
c (Å)
c (I4/mmm)
c (Immm)
Tet.
Tet. + Orth.
Orth.
0.0 0.2 0.4 0.6 0.8 1.0 1.2 1.4
x_nom
(b)
Sr2VO3.92 (x_nom = 0.0)
I4/mmm
a = 3.8381
c = 12.5551
R_wp = 4.95
Sr2VO3.62H0.38 (x_nom = 0.5)
I4/mmm
a = 3.8171
c = 12.6711
R_wp = 6.17
Sr2VO2.99H1.01 (x_nom = 1.2)
Immm
a = 3.6811, b = 3.8878
c = 12.7839
R_wp = 3.47
S (Q)
Q (Å^-1)

Fig.1

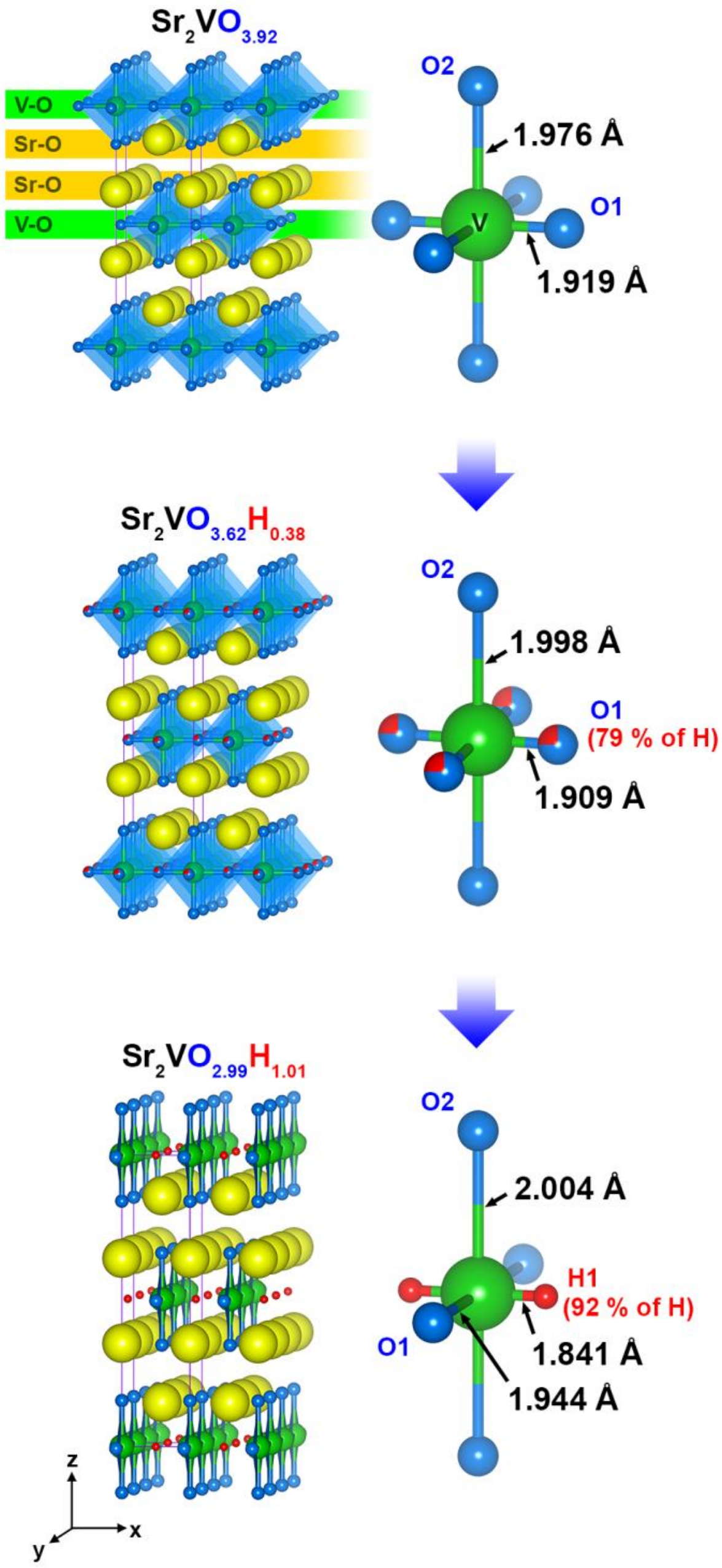

Fig.2

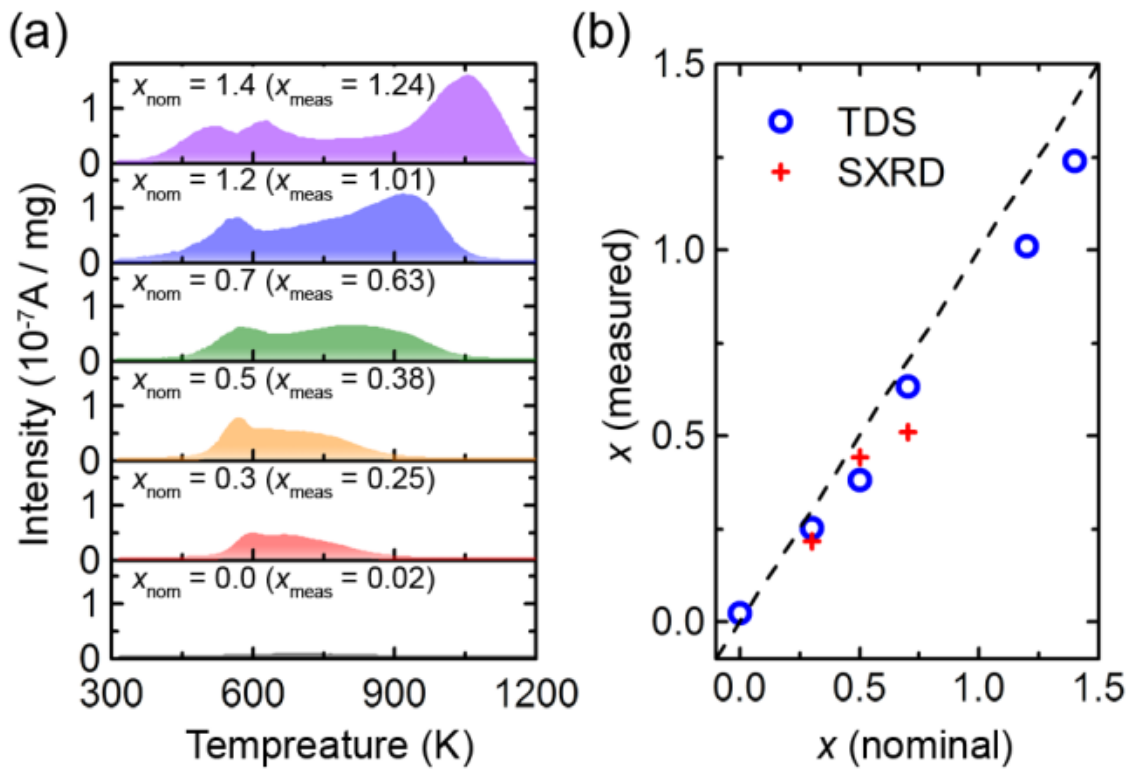

Fig.3

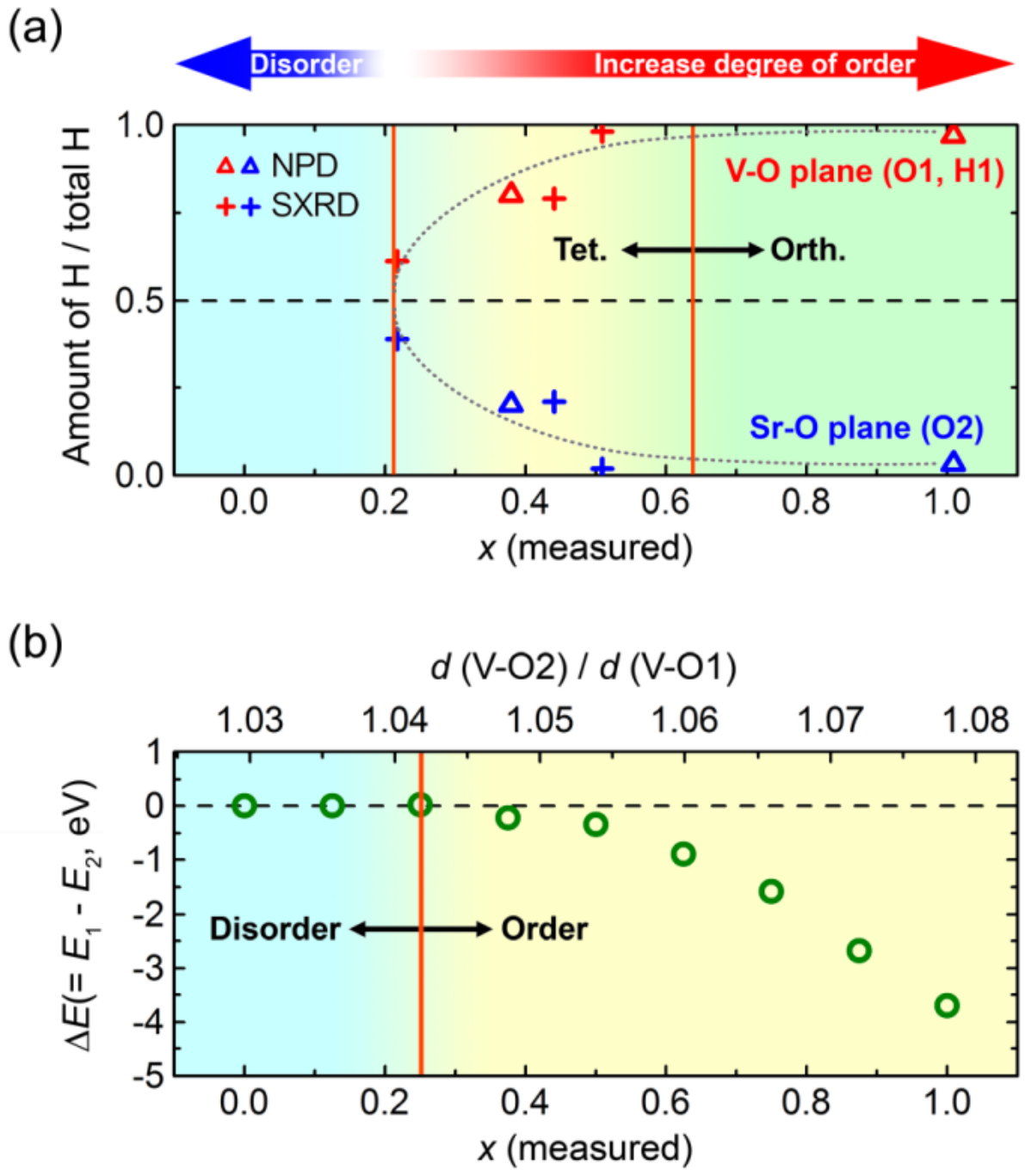

(a)
Disorder
Increase degree of order
NPD
SXRD
V-O plane (O1, H1)
Tet.
Orth.
Sr-O plane (O2)
Amount of H / total H
x (measured)
(b)
d (V-O2) / d (V-O1)
Disorder
Order
ΔE(= E1 - E2, eV)
x (measured)

Fig.4

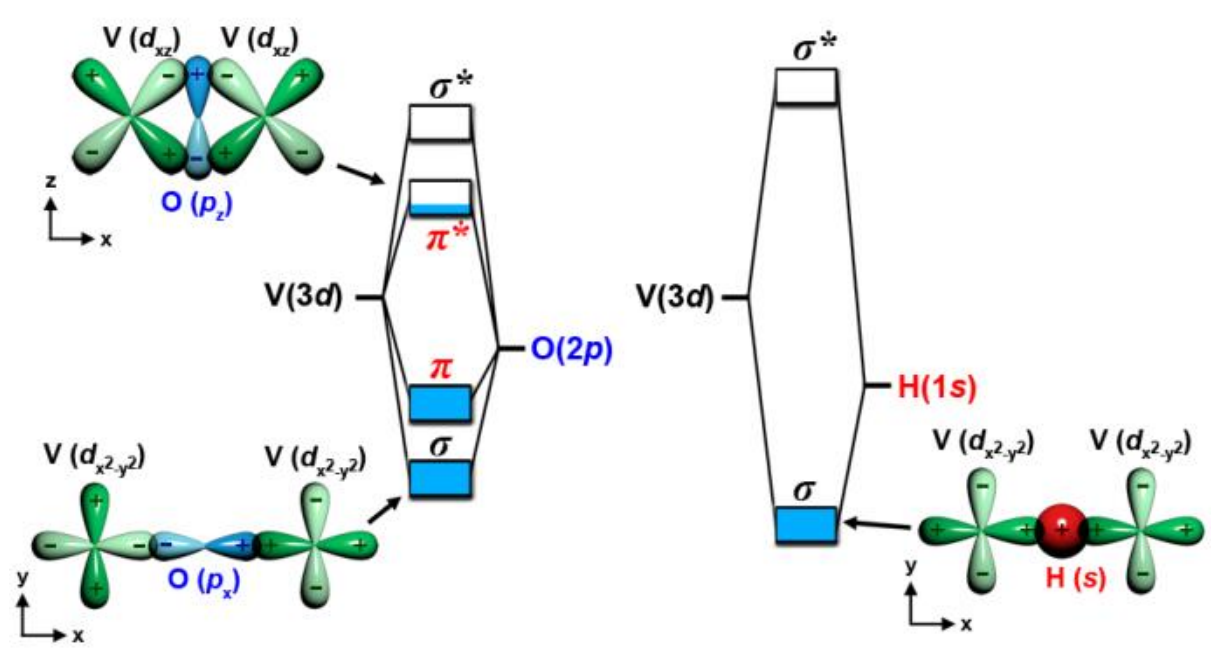

Fig.5

## Supplemental Information

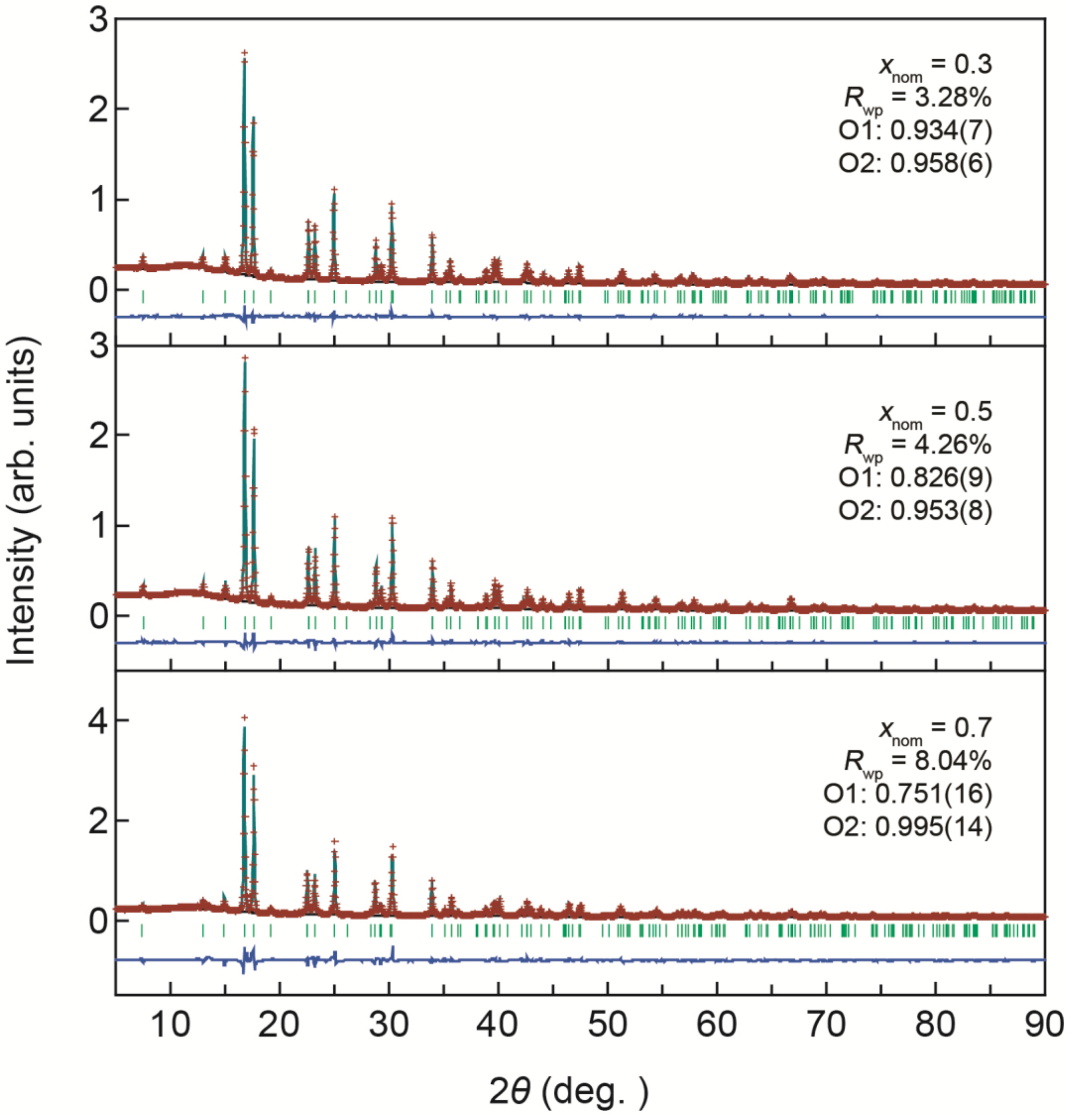

**Figure S1.** Rietveld fitting of synchrotron diffraction for $Sr_2VO_{4-x}H_x$ ($x_{nom}$ = 0.3, 0.5, and 0.7, λ = 0.82722 Å).

**Table S1.** Lattice parameters of $Sr_2VO_{4-x}H_x$ ($x_{nom}$=0.0, 0.5, and 1.2) carried out by Rietveld fitting of NPD data.

| Atom | Site | Occ. | *x* | *y* | *z* | $B_{iso}$(Å$^2$) |
|---|---|---|---|---|---|---|
| **(a) $Sr_2VO_{3.92}$ (*I*4/*mmm*, *Z* = 2)** | | | | | | |
| ***a* = 3.838083(78) Å, *c* = 12.555137(907) Å, *V* = 184.9483 Å$^3$, $R_{wp}$ = 4.95** | | | | | | |
| **$Sr_2VO_{4-x}H_x$ = 93.59%, SrO = 6.41%** | | | | | | |
| **Sr** | 4e | 1 | 0 | 0 | 0.355868(101) | 0.547(28) |
| **V** | 2a | 1 | 0 | 0 | 0 | 0.237 |
| **O1** | 4c | 0.989(4) | 0 | 0.5 | 0 | 0.493(28) |
| **O2** | 4e | 0.969(4) | 0 | 0 | 0.157426(110) | 0.434(35) |
| **(b) $Sr_2VO_{3.62}H_{0.38}$ (*I*4/*mmm*, *Z* = 2)** | | | | | | |
| ***a* = 3.817189(83) Å, *c* = 12.671137(823) Å, *V* = 184.6303 Å$^3$, $R_{wp}$ = 6.17** | | | | | | |
| **$Sr_2VO_{4-x}H_x$ = 81.07%, SrO = 18.93%** | | | | | | |
| **Sr** | 4e | 1 | 0 | 0 | 0.353835(112) | 0.550(22) |
| **V** | 2a | 1 | 0 | 0 | 0 | 0.237 |
| **O1** | 4c | O: 0.850<br>H: 0.150 | 0 | 0.5 | 0 | 0.398(22) |
| **O2** | 4e | O: 0.960<br>H: 0.040 | 0 | 0 | 0.157719(125) | 0.398(22) |
| **(c) $Sr_2VO_{2.99}H_{1.01}$ (*Immm*, *Z* = 2)** | | | | | | |
| ***a* = 3.681063(128) Å, *b* = 3.887802(117) Å, *c* = 12.783869(353) Å, *V* = 182.9531 Å$^3$, $R_{wp}$ = 3.47** | | | | | | |
| **$Sr_2VO_{4-x}H_x$ = 92.77%, SrO = 7.23%** | | | | | | |
| **Sr** | 4i | 1 | 0 | 0 | 0.351393(95) | 0.495(22) |
| **V** | 2a | 1 | 0 | 0 | 0 | 0.237 |
| **O1** | 2d | O: 0.950<br>H: 0.050 | 0 | 0.5 | 0 | 0.183(37) |
| **O2** | 4i | O: 0.985<br>H: 0.015 | 0 | 0 | 0.156735(125) | 0.378(26) |
| **H1** | 2b | O: 0.070<br>H: 0.930 | 0.5 | 0 | 0 | 1.693(95) |

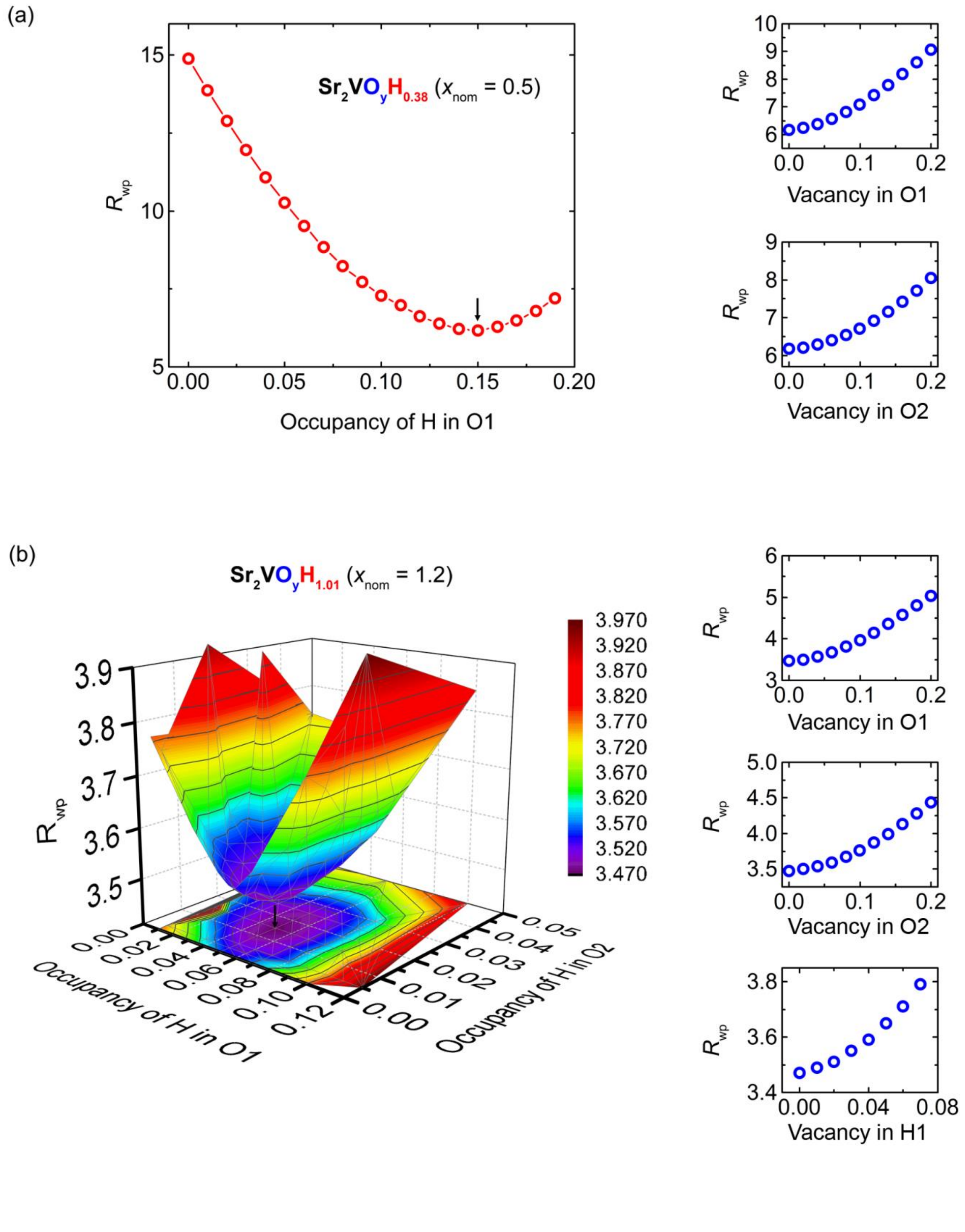

**Figure S2.** Residual factor of refinement $R_{wp}$ for NPD data of $Sr_2VO_{4-x}H_x$ for (a) $x_{nom}$ = 0.5 and (b) $x_{nom}$ = 1.2. Total amount of hydrogen is fixed to 0.38 and 1.01 for each pattern based on the TDS data. Right-hand side: $R_{wp}$ for refinement of NPD patterns as a function of site occupancy for the each anion site. In this analysis, atomic parameters for each anion site, i.e., (x, y, z), $B_{iso}$, and hydrogen amount were fixed, whereas the site occupancy for the oxygen was treated as a parameter.